\documentclass[a4paper,conference]{IEEEtran}
\IEEEoverridecommandlockouts
\usepackage{cite}
\usepackage{amsmath,amssymb,amsfonts}
\usepackage{algorithmic}
\usepackage{graphicx}
\usepackage{textcomp}
\usepackage{xcolor}
\usepackage[hidelinks]{hyperref} 
\usepackage{cleveref} 
\usepackage{float}
\usepackage{enumitem}
\usepackage{esvect}

\usepackage{tabu}
\usepackage{booktabs}
\usepackage{tikz}
\usepackage{textpos}
\usepackage{framed}

 \usepackage{makecell}
 \usepackage{multirow}
 \usepackage{adjustbox}
 \usepackage{subcaption}
 \usepackage{enumitem}
 \usepackage{comment}
 \usepackage{flushend}

\begin{document}

         \title{Towards Credential-based Device Registration \\in DApps for DePINs with ZKPs}



        \author{\IEEEauthorblockN{Jonathan Heiss}
\IEEEauthorblockA{\textit{Information Systems Engineering} \\
\textit{TU Berlin}, Germany \\
jh@ise.tu-berlin.de}
\and
\IEEEauthorblockN{Fernando Castillo}
\IEEEauthorblockA{\textit{Information Systems Engineering}\\
\textit{TU Berlin}, Germany \\
fc@ise.tu-berlin.de}
\and
\IEEEauthorblockN{Xinxin Fan}
\IEEEauthorblockA{\textit{IoTeX} \\
Menlo Park, CA 94025, USA \\
xinxin@iotex.io}
}

	\maketitle

	\begin{abstract}
        Decentralized Physical Infrastructure Networks (DePINS) are secured and governed by blockchains but beyond crypto-economic incentives, they lack measures to establish trust in participating devices and their services. 
        The verification of relevant device credentials during device registration helps to overcome this problem. 
        However, on-chain verification in decentralized applications (dApp) discloses potentially confidential device attributes whereas off-chain verification introduces undesirable trust assumptions. 
        In this paper, we propose a \textit{credential-based device registration} (CDR) mechanism that verifies device credentials on the blockchain and leverages zero-knowledge proofs (ZKP) to protect confidential device attributes from being disclosed. 
        We characterize CDR for DePINs, present a general system model, and technically evaluate CDR using zkSNARKs with Groth16~\cite{groth2016size} and Marlin~\cite{chiesa2020marlin}. 
        Our experiments give first insights into performance impacts and reveal a tradeoff between the applied proof systems. 
	\end{abstract}


	\section{Introduction}
	\label{sec:introduction}
	In untrusted environments like the Internet of Things (IoT), blockchains have manifested as a solution for managing devices and their data without introducing undesirable third-party dependencies or trust assumptions. 
Examples of such decentralized applications (dApps) include netting from smart meter data in energy grids~\cite{peise2021blockchain,eberhardt2020privacy}, product tracing through sensor measurements in supply chains~\cite{sund2020supplychain,heiss2023verifiable}, and decentralized federated learning on healthcare data collected by wearables~\cite{heiss2022advancing,lee2024end}.

More recently, Decentralized Physical Infrastructure Networks (DePINs)~\cite{ballandies2023taxonomy} have emerged as a class of dApps that add crypto-economic mechanisms to the dApp's smart contracts to incentivize the provisioning and consumption of device-enabled services. 
This has led to the formation of large decentralized networks of devices that collectively offer project-specific services, providing viable alternatives to centralized service models. 
Examples of industrial projects utilizing DePINs include IoTeX for sensing services~\cite{misc:iotex}, Helium for connectivity services~\cite{misc:helium}, StreamR for data streaming services~\cite{misc:streamr}, and Acurast for computational services~\cite{misc:accurast}.

In DePINs, token-based incentive schemes represent the key element to governing and establishing trust in the DePIN’s service model.
Such schemes rely heavily on off-chain data from the devices to trigger token issuance on completed service provisioning. 
However, the reliance on off-chain data represents a security threat. 
Since blockchain security guarantees do not extend beyond the smart contract's application logic, malicious actors can corrupt data to trigger unwarranted token issuance.
This makes \textit{trustworthy data and service provisioning} indispensable for the success of DePINs.

\textit{Trustworthy data provisioning} is challenging as devices typically do not communicate directly with smart contracts but data provisioning is intermediated by device owners or third-party oracles~\cite{heiss2019oracles}. 
To prevent data corruption authenticity proofs created by the devices are verified on the blockchain using the devices’ public keys. 
Additionally, zero-knowledge proofs (ZKP) can be employed for trustworthy pre-processing~\cite{park2020smart,wan2022zk,heiss2023trustworthy,heiss2021trustworthy} enabling end-to-end verification of \textit{data in use} between devices and smart contracts. 

The \textit{trustworthiness of the service provisioning} is hard to guarantee without measures to verify or enforce service qualities. 
In DePINs, such qualities strongly depend on the capabilities and attributes of the devices providing specific service types.
Service quality assurance, consequently, requires devices to meet certain attribute-based criteria.
Computation services may require a minimum of CPU cycles or memory capacity, connectivity services may require a certain level of bandwidth or range, and sensing services may require specific sensor capabilities or devices to be placed in certain locations.

Transparently enforcing such conditions on device attributes in decentralized settings is challenging and requires an appropriate model for managing device identity attributes.
For that, concepts of the Self-Sovereign Identity (SSI) paradigm represent a promising approach promoting decentralized and user-centric management of identity-related data.
Applied to DePINs, device manufacturers and other identity providers can issue verifiable credentials (VCs)~\cite{misc:W3_VerifCred} that attest to device attributes, which devices or their owners can present to third-party consumers.

A problem for decentralized attribute verification is that device credentials often contain confidential information that must not be made public, like private residence locations or security-relevant parameters like software versions.
Blockchain-based verification of signature-based attestations guarantees transparency but publicly exposes these attributes violating confidentiality requirements.
Alternatively, off-chain verification which is the common practice in many DePIN projects introduces undesirable trust assumptions and risks excluding eligible devices or registering non-eligible ones, thereby undermining promised decentralization.

Addressing this conflict of confidentiality and transparency, we propose a mechanism for \textit{credential-based device registration} (CDR) for dApps in DePINs that enables non-disclosing verification of device credentials on the blockchain using ZKPs. 
Verified devices are registered on-chain with their public key which is later used to validate the authenticity of data received from the devices. 
The mechanism builds upon and extends the W3C verifiable credential model~\cite{misc:W3_VerifCred} that best fits the contextual demands of dApps in DePINs.
In this preliminary work, we make three individual contributions:
\begin{itemize}
    \item We present a system model for DePINs that supports integration with the W3C VC model and the proposed CDR mechanism. The model is underpinned with examples of registration conditions, device attributes, and VC issuers facilitating its instantiation. 

    \item We present a device registration mechanism that leverages verifiable device credentials to transparently decide registration conditions on the blockchain. To hide confidential device attributes, ZKPs are used for off-chain pre-processing yielding non-disclosing and on-chain verifiable proofs of registration conditions. 
        
    \item We evaluate the system through a prototypical implementation using ZoKrates~\cite{eberhardt2018zokrates} for zkSNARK creation and verification on Ethereum~\cite{wood2014ethereum}. We conduct initial experiments on test credentials with Groth16~\cite{groth2016size} and Marlin~\cite{chiesa2020marlin}. The results give first insights into the method's performance behavior. 
\end{itemize}

	\section{Preliminaries}
	\label{sec:preliminaries}
	As central concepts, this work relies on \textit{verifiable credentials} (VC) and \textit{zero-knowledge proofs} (ZKP).

\subsection{Verifiable Credentials}
\label{sec:verifiable_credentials}
\begin{figure}[h] %
	\centering
	\includegraphics[width=0.9\columnwidth]{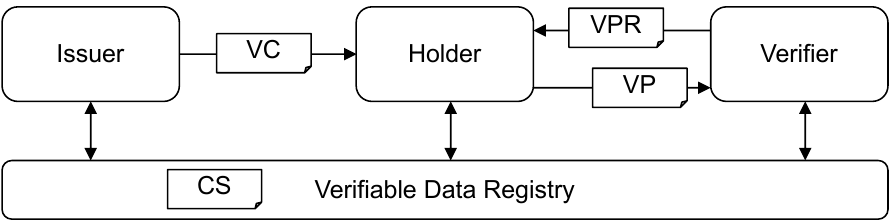}
	\caption{Verifiable Credential Model based on~\cite{misc:W3_VerifCred}}
	\label{fig:vc_model}
\end{figure}

The W3C recommendation for VC~\cite{misc:W3_VerifCred} advocates for a user-centric identity management framework. 
As depicted in Figure~\ref{fig:vc_model}, identity attribute claims are issued by a trusted \textit{issuer} as VCs to the \textit{holder}, who then securely stores them, e.g., in a digital wallet.
A \textit{credential} (CR) contains claims of an issuer about the device that consists of a 3-tuple comprising subject, attribute, and value, e.g., $(\text{Alice}, \text{lives\_in}, \text{NYC})$.
A VC adds authorship of issuers through attestations to a credential's claims that can be cryptographically verified.
The holder can then independently present a selection of these verifiable attribute claims as a \emph{Verifiable Presentation}~(VP) to a \emph{verifier}.
VPs can be created with ZKPs to hide confidential information from the verifier.  
Verifiers can specify the VP through a \textit{Verifiable Presentation Request}~(VPR)~\cite{misc:W3_VPR} according to the needs of the application at hand. 
Unlike the focus of this paper, the VC model assumes the verifier operates off the blockchain.
Blockchains are utilized solely to implement \emph{Verifiable Data Registries}~(VDR), which store public artifacts such as identifiers, public keys, or \emph{credential schemas}~(CS) that define the structure and verification process of VCs.

\subsection{Non-Interactive Zero-knowledge Proofs}
\label{sec:zkp}
Non-Interactive Zero-knowledge Proofs (NIZK) allow a prover to convince a verifier of a statement in one message (non-interactive) and without disclosing any confidential information despite the statement itself. 
Zero-knowledge succinct non-interactive arguments of knowledge (zkSNARKs) represent a specific category of NIZK known for their compact proof sizes and efficient verification times. 
The zkSNARKs procedure can be conceptualized in three main operations: 

\begin{itemize}
\item The \textit{setup} ($setup(ecs, srs) \rightarrow (PK, VK)$) generates a public asymmetric key pair derived from the executable constraint systems ($ecs$) and structured reference string ($srs$). 
Both proving and verification keys ($PK, VK$) are tied to the $ecs$ which encodes the program logic in a provable representation.
This process assumes a secure disposal of the $srs$ to prevent the creation of fake proofs.

\item The \textit{proving} ($P(ecs, x, x', w, PK) \rightarrow \pi$) occurs in two steps: 
Firstly, a witness $w$ is created by executing the $ecs$ on the public inputs $x$ and the private inputs $x'$ constituting the proof arguments. 
The witness $w$ signifies a valid variable assignment for the $ecs$ inputs. 
Subsequently, the proof $\pi$ is generated from the witness using the $PK$.

\item The \textit{verification} ($V(\pi, x, VK) \rightarrow \{0, 1\}$) evaluates the proof $\pi$ and the public inputs $x$ using the verification key $VK$. 
\end{itemize}

ZkSNARKs facilitates Verifiable Off-Chain Computing~\cite{eberhardt2018off} (VOC) which helps to overcome the privacy and scalability limitations of blockchains by offloading computation without compromising the blockchain's integrity.
VOC is technically supported by \textit{ZoKrates}~\cite{eberhardt2018zokrates}, a language and toolbox that facilitates the development of zkSNARKs-based VOC for the Ethereum blockchain~\cite{wood2014ethereum}.

        \section{Model}
	\label{sec:04_model}
	\begin{figure*}[t]
    \centering
    \includegraphics[width=0.95\textwidth]{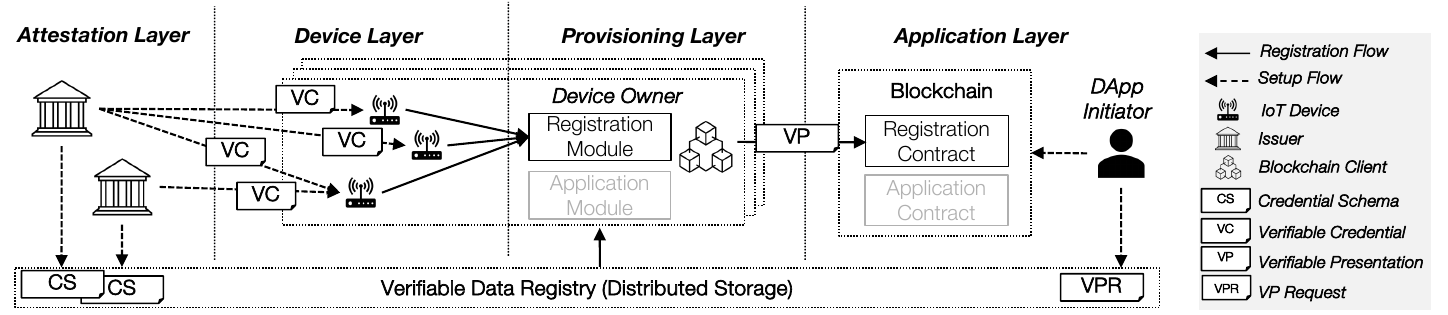}
    \caption{System Model for DApps in DePINs Enabling Device Registration with Verifiable Credentials}
    \label{fig:model}
\end{figure*}

To set the scenes, in this section, we first characterize credential-based device registration (CDR) through examples of device attributes, registration conditions, and credential issuers, then we present a general system model for CDR in dApps, and finally outline the threat model and associated objectives for designing a CDR mechanism in DePINs.

\subsection{Characterizing Credential-based Device Registration}
\label{sec:CDR}
We characterize CDRs by giving examples of \textit{conditional checks} executed on \textit{device attributes} attested to by \textit{credential issuers}.

\subsubsection{Conditional Checks}
\label{sec:conditions}
Registration conditions can be realized with equality, range, membership, and time-dependent proofs~\cite{heiss2022non}.
\begin{itemize}
    \item \textit{Equality checks} are used in dApps if a device attribute must be equal to a predefined value.  

    \item \textit{Range checks} are used in DApps if a numeric attribute is required to be in a specific range indicated through an upper and/or lower bound. 
    
    \item \textit{Membership checks} are used in dApps if it is required that a device's attribute value $val$ is in a predefined finite set $S = {s1, s2, ..., sn}$ such that $val \in S$. 
    
    \item \textit{Relative time-dependent checks} are used in dApps if a date- or time-based attribute is required to be in a range that has boundaries relative to the current date or a timestamp.
\end{itemize}

\subsubsection{Device Attributes} 
\label{sec:device_attributes}
Such registration conditions rely on device attributes of different types. The following examples, build upon and extend the collection presented in~\cite{dayaratne2024ssi4iot}.

\begin{itemize}
    \item \textit{Identity attributes} can be categorized in \textit{static attributes}, e.g., public key, serial number, manufactured date; and \textit{dynamic attributes}, e.g., firmware version, last\_updated\_on, owner\_id. 
    
    \item \textit{Capability attributes} describe available resources and functionality of the devices regarding \textit{communication}, e.g., wired, wireless, or satellite; \textit{computation}, e.g., clock speed, number of cores; \textit{memory}, e.g., size.
    
    \item \textit{Configuration attributes} can comprise the device \textit{thresholds}, e.g., max and min values; \textit{security parameters}, e.g., elliptic curve types or hash algorithm; \textit{communication types}, e.g., scheduling of cron job intervals. 
    
    \item \textit{Installation attributes} further describe the installation or onboarding process. This may concern the \textit{installer}, e.g., installer\_id, certificate, association; and the \textit{installation}, e.g., time, location, sealing. 
\end{itemize}

\subsubsection{VC Issuers} 
\label{sec:vc_issuers}
Device attributes can be attested to by different types of issuers as presented in~\cite{dayaratne2024ssi4iot}.
\begin{itemize}
    \item \textit{Manufacturers} have access to critical device information and control large parts of the device's lifecycle which allows for attesting to many attributes. 
    
    \item \textit{Regulators} can establish and enforce industry standards and regulations, e.g., for medical devices or smart meters.  
    
    \item \textit{Service providers} can attest to attributes associated with their offered service, e.g., device installation or firmware updates. 
    
    \item \textit{Device owners} can attest to a variety of attributes including self-attesting the device ownership. 
\end{itemize}

\subsection{System Model}
\label{sec:system_overview}
While previous examples are intended to help design CDR conditions, their integration with the W3C VC model and DePIN applications can be challenging. 
For that, we propose a general system model for device registration in dApps that integrates the W3C VC model presented in Section~\ref{sec:verifiable_credentials}. 
As depicted in Figure~\ref{fig:model}, the system consists of four layers.

\subsubsection{Attestation Layer}
On the attestation layer, issuers attest to the attributes of the device. 
Device attributes can be attested to by different issuers. 
For example, the same industry certificate may be issued by different accredited service providers or authorities. 
The resulting verifiable credentials are provisioned to the devices where they are protected against unauthorized access. 
The issuer uses the \textit{verifiable data registry} (VDR) to publish the corresponding \textit{credential schema} (CS).

\subsubsection{Device Layer}
On the device layer, devices collect data from the service provisioning relevant to the smart contract-based application logic, e.g., token issuance in DePINs. 
Devices are characterized by attributes that are subject to the CDR condition.
Such attributes are attested to by issuers and contained in a \textit{verifiable credential} (VCs) which can only be accessed by the device owner and the associated issuers. 
To include resource constraint devices, we assume separate the registration and application logic as well as the interaction with the smart contracts to the device owners who are assumed to act on behalf of the devices.

\subsubsection{Provisioning Layer}
On the provisioning layer, device owners provision data obtained from the devices to the smart contracts through a \textit{blockchain client} using their blockchain account address. 
The \textit{application module} (AM) is responsible for tasks related to data provisioning once a device is registered which may include a pre-processing of sensor data~\cite{heiss2021trustworthy} or the creation of a service provisioning proof. %
The \textit{registration module} (RM) is used for device registration and the focus of this work. 
Here, the \textit{verifiable presentation} (VP) is created from the devices' VC according to the \textit{verifiable presentation request} (VPR).

\subsubsection{Application Layer}
The application layer consists of two types of smart contracts running on a blockchain infrastructure. 
They are deployed by the \textit{dApp initiator} who represents the project provider in DePIN projects. 
\textit{Application contracts} implement context-specific application logic that relies on the off-chain device data, e.g., the token issuance. 
\textit{Registration contracts} check the CDR condition that devices must satisfy to be accepted as data sources for the application logic. 
The registration contract takes the VP as input which is created by the device owner according to the VPR specified by the initiator.

\subsection{Threat Model and Challenges}
In the previous model, we expect attacks from device owners and the dApp initiator whereas issuers are assumed to be trusted, following the W3C VC model~\cite{misc:W3_VerifCred}.
Device owners may want to corrupt the registration mechanism by submitting VPs containing false claims to obtain tokens from services provided by non-eligible devices. 
Where possible, the dApp initiator may try to register non-eligible devices acting as device owners. 
Both, device owners and the dApp initiator may also try to obtain access to confidential device attributes. 
To prevent such threats, we derive the following objectives for a CDR mechanism in the described system model:
\begin{itemize}
    \item \textit{Verifiability}: The device registration must be verifiable by all device owners involved in and affected by the dApp. This includes the correctness of the issuers' attestations and the validation of the registration condition. 
    \item \textit{Confidentiality}: Device attributes must not be disclosed to anyone but the associated device owner. This means that neither the attestations nor the registration condition can be verified directly by other device owners or be verified on the blockchain. 
\end{itemize}

        \section{System Design}
	\label{sec:05_provingService}
\begin{figure}[t]
    \centering
    \includegraphics[width=1\columnwidth]{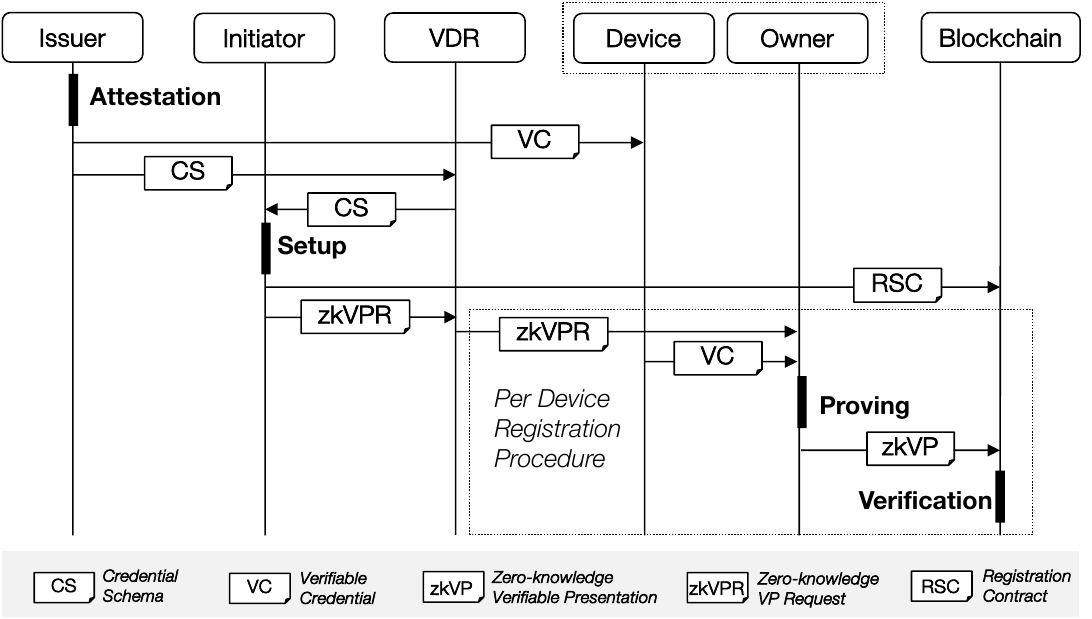}
    \caption{Device Registration Procedure}
    \label{fig:procedure}
\end{figure}

	To achieve previously formulated objectives, in this section, we propose a mechanism for credential-based device registration (CDR) that leverages zkSNARKs to make the off-chain validation of CDR conditions verifiable on the blockchain without disclosing confidential device attributes. 
Distinctive artifacts are the \textit{zero-knowledge verifiable presentation request} (zkVPR) created by the initiator that helps the device owner to create a \textit{zero-knowledge verifiable presentation} (zkVP) which is validated by the registration contract as a non-disclosing registration request. 
On successful verification of the zkVP, the device's public key ($pubk_d$) is registered on-chain and can later be used to authenticate device-generated data submitted to the dApp. 
The procedure is depicted in Figure~\ref{fig:procedure} and consists of three phases. 
\begin{enumerate}
    \item [A.] The \textit{attestation} is considered a pre-requisite executed by the issuers who create the VCs by signing the device attributes.
    \item [B.] The \textit{setup} is executed once per application by the initiator who creates and deploys the zkVPR and the registration contract.
    \item [C.] The \textit{registration} is executed for each device. It consists of a proving where the device owner creates the zkVP and the verification where the zkVP is verified by the registration contract. 
\end{enumerate}

In the following, we describe each phase in more detail. 

\subsection{Attestation}
\label{sec:attestation}
During the attestation, the issuer creates a verifiable credential~(VC) by signing each claim of a credential (CR) individually with the \emph{issuer secret key} $seck_i$: $Attest(CR, seck_i) \rightarrow (VC)$.
This results in a VC consisting of a set of \emph{verifiable claims}~($vcl$) ($VC = \{vcl_1,...,vcl_n\}$) each represented as a claim-signature pair ($vcl = \{cl, sig\}$).

We assume that each device has a device-specific public key ($pubk_d$) that could be attested to by a suitable issuer, e.g., a public key authority. 
The $pubk_d$ will be used on-chain to verify authenticity proofs of the device created with the corresponding device secret key ($seck_d$).
We treat the certificate as a VCL that contains $pubk_d$ as the attribute. 

The resulting VC is securely stored on the device and the corresponding CS is published on the \textit{Verifiable Data Registry} (VDR), a distributed storage system.

\subsection{Setup}
\label{sec:initialization}
For the setup, the initiator creates and deploys the \textit{zkVPR} and the \textit{registration contract}. 
A zkVPR consists of the \textit{zero-knowledge proof specification} ($zkSpec$), the \textit{proving key} ($pk$), and a reference to the corresponding CS. Additional meta information can be added. 
To deploy the registration contract, the \textit{verification key} ($vk$) is required to validate the correctness of the zkVP. 
The setup can be described in three steps. 

\subsubsection{Proof Specification}
The zkSpec should allow the device owners to create the zkVP and, for that, it specifies the logic of the CDR condition to be proven in zero-knowledge and the required inputs.
CDR conditions typically consist of two operations, both executed for each relevant VCL:

\begin{itemize}
    \item \textit{Authenticity Check}: The issuer's signature ($sig$) is verified using the issuer's public key ($pk_i$): $Verify(sig, cl, pk_i) \rightarrow \{1,0\}$
    \item \textit{Conditional Check}: As described in Section~\ref{sec:conditions}, an attribute-specific registration condition is checked using some auxiliary data ($aux$) like thresholds for range proofs or candidate lists for membership proofs: $Check(att, aux) \rightarrow \{1,0\}$
\end{itemize}
Public inputs ($x$) are $pk_i$ and $aux$ whereas private inputs ($x'$) are $sig$, $cl$, and $att$.
While the composition of VCLs can vary per application, each zkSpec of a CDR condition must contain the device's public key ($pubk_d$) as a public input to bind the zkVP and its on-chain validation to the device. 
The $pubk_d$ is later added to the device registry of the registration contract. 
If both checks pass for all VCLs, the device's public key is returned to bind the device attributes to the key. 

\subsubsection{Key Generation}
The initiator creates the zero-knowledge \textit{proving key} ($pk$) and the \textit{verification key} ($vk$).
For that, first, the zkSpec is compiled into an executable constraint system ($ecs$) to enable the assertion of computational correctness through a zkSNARK. 
Using the $ecs$, the device owner generates $pk$ and $vk$ as described in Section~\ref{sec:zkp}: $KeyGen(ecs,srs) \rightarrow (pk,vk)$. 
The keys are bound to the $ecs$ and enable the device owner to create a verifiable $zkSpec$-specific zkVP with the $pk$ that can be verified by the registration contract with the corresponding $vk$.

\subsubsection{Deployment}
The initiator publishes the $zkSpec$, the $pk$, and the CS-reference as zkVPR to the VDR where it becomes accessible by the device owners.
The zkVPR-reference and the $vk$ are integrated into the registration contract which is finally deployed to the blockchain. 

\subsection{Registration}
\label{sec:registration}
The registration consists of the off-chain proving by the device owner and the on-chain verification by the registration contract.

\subsubsection{Proving}
\label{sec:proving}
The device owner first collects the necessary information and artifacts, that is, the zkVPR-reference from the registration contract and the zkVPR and CS from the VDR (the CS-reference is contained in the zkVPR). 
The \textit{zkSpec} and the CS provide the necessary information to create the required zkVP. 
Accordingly, the device owner accesses the requested verifiable claims ($vcl$) from the device's secure storage which represents the private inputs ($x'$) to the proving. 

To create the zkVP, the device owner re-compiles the zkSpec into a $ecs$ and creates the zkVP in two steps as described in Section~\ref{sec:zkp}.
First, the witness is generated by executing the $ecs$ using the $vcl$ as private inputs $x'$ and the auxiliary data $aux$ and public keys ($pubk_d, pubk_i$) as public inputs $x$. 
Second, the zkVP is created on the witness $w$ using the $pk$ contained in the zkVPR. 
A successful execution results in the zkVP consisting of $pubk_d$, $aux$, and the proof of computational correctness $\pi$. 

\subsubsection{Verification}
\label{sec:verification}
The verification is executed by the registration contract on the zkVP obtained from the device owners as a registration request. 
The registration contract implements (1) the device registry, and functionality for (2) the ZKP verification, and (3) additional checks in the public inputs ($x$). 
It is executed in two steps: 
\begin{itemize}
    \item \textit{Proof Verification}: The zkVP's correctness is validated using the verification key, public inputs, and proof: $Verify(\pi, vk, pubk_i, pubk_d, aux) \rightarrow \{1,0\}$.
    \item \textit{Input Verification}: The applied public inputs ($pubk_i, pubk_d, aux$) are checked against a predefined list of inputs stored in the registration contract. 
\end{itemize}

If both checks pass, the device's public key is added to the device registry.
With that, it can be enforced that only registered devices can call functions in the application contract. 
As a provisioning condition, devices must create a signature over the device-generated data using the secret key that matches a public key in the registry.

	\section{Evaluation}
	\label{sec:evaluation}
	Given a detailed specification of the credential-based registration (CDR) mechanism, in this section, we describe our prototypical implementation and present our initial experiments.

\subsection{Implementation}
To demonstrate the technical feasibility of our proposal, we prototypically implement the CDR mechanism.
%
The \textit{attestation} is realized with a Python script that creates EdDSA signatures over the test credentials used for the experimentation. 
It should be noted that the elliptic curve applied for signature creation must be supported by the proof system and the verification environment. 
Respecting these dependencies, we use the babyjubjub curve (ALT\_BN128) that is supported by the Ethereum Virtual Machine~\cite{wood2014ethereum} and, hence, allows for verification of babyjubjub-based SNARKs. 
%
The \textit{setup} and \textit{proving} are realized with ZoKrates~\cite{eberhardt2018zokrates}. 
We implement the proof specification in the ZoKrates DSL resulting in a human-readable and small-sized artifact that is suited for efficient sharing and allows the device owner to understand and double-check the proof logic. 
Furthermore, we use the ZoKrates Command Line Interface for the compilation into a $ecs$, key generation, witness computation, and proof generation. 
%
For \textit{verification}, we use the Solidity verifier smart contract generated by ZoKrates. 
It implements the routines required to verify the ZKP in the verytx() function using the integrated verification key. 
Smart contracts are hosted on a locally simulated Ethereum blockchain using Hardhat\footnote{https://hardhat.org/hardhat-runner/docs/getting-started} test suit.

\begin{table*}[t]
\centering
\caption{Experimental Results}
\begin{tabular}{lrrrrrrrr}
\toprule
    \bf Proof & 
    \bf Scheme & 
    \bf TX Cost (Gas) & 
    \bf Witness (s) & 
    \bf Setup (s) & 
    \bf Proof (s) & 
    \bf Compiled (MB) & 
    \bf PK (MB) & 
    \bf VK (KB) \\
\midrule

Range & Groth16 & 595\,k & 3 & 3 & 3 & 400 & 59 & 8 \\
Membership & Groth16 & 623\,k & 5 & 7 & 6 & 752 & 126 & 8 \\
Equality & Groth16 & 495\,k & 2 & 3 & 3 & 400 & 59 & 8 \\

Range & Marlin & 1007\,k & 2 & 1322 & 39 & 453 & 2322 & 9 \\
Membership & Marlin & 1035\,k & 4 & 2674 & 79 & 906 & 4624 & 9 \\
Equality & Marlin & 889\,k & 2 & 1338 & 38 & 453 & 2322 & 9 \\

\bottomrule
\end{tabular}
\label{table:measurements}
\end{table*}

\subsection{Experimentation}
To obtain insights into the practicality of the system, we conduct initial experiments on our prototypical implementation.

\subsubsection{Objective}
The usage of zkSNARKs and blockchains adds considerable performance overhead to the (zk)VPR creation, the (zk)VP creation, and the (zk)VP verification. 
Such overheads may lead to unacceptable resource requirements, execution times, or transaction costs preventing adoption in practical settings.
The objective of these experiments is to get initial insights into how the usage of zkSNARKs for anonymous device credentials in dApps negatively impacts such performance-related qualities.

\subsubsection{Design}
To achieve this objective, we implement several zero-knowledge proof specifications (zkSpec) using different registration conditions, compile them into zkVPRs, execute them on typical device attributes, and verify the resulting zkVPs in an Ethereum Virtual Machines~\cite{wood2014ethereum}.
zkSpecs define the (1) validation of the issuer's signature and (2) an attribute-based condition for each device attribute.
In the experiments, we use three typical conditions as described in Section~\ref{sec:conditions}:
 
\begin{itemize}
    \item First, we use a \textit{range check} to validate that only devices are registered that have firmware with a minimum version number. The private device attribute is the firmware number and the public threshold is the minimum version. 
    \item Second, we use a \textit{membership check} to validate that only devices within a certain regional range are registered. The private device attribute is a postcode representing the device's location and the range is defined by a public list of permissible postcodes. 
    \item Third, we use an \textit{equality check} to validate that only devices with specific measurement types are registered. The device attribute is a code representing the device's measurement type. It is checked against a predefined type. 
\end{itemize}

In anticipation of a tradeoff between performance and security, we use two different proof systems to technically realize zkSNARKs: First, we use \textit{Groth16}~\cite{groth2016size} which is well-established and known to be efficient but relies on a trusted setup as described in Section~\ref{sec:zkp}. Second, we use \textit{Marlin}~\cite{chiesa2020marlin} which mitigates trust assumptions from the setup through a universal and updatable structured reference string ($srs$) which, however, is expected to cause a performance loss.

We execute each experiment using test credentials on a MacBook Pro (Model Identifier: Mac14,10, Model Number: MNW83CI/A) with an Apple M2 Pro chip (12 cores: 8 performance and 4 efficiency), 16 GB of memory, and System Firmware Version 10151.101.3. 
For each proof and scheme combination, we measure the transaction cost in terms of Gas, the time taken to generate the witness, setup time, and proof generation time in seconds. 
Additionally, we report the sizes of the compiled $ecs$ and proving key (PK) in megabytes (MB), and verification key (VK) in kilobytes (KB).

\subsubsection{Results}
As depicted in Table~\ref{table:measurements}, the experimental results show the trade-offs between the proof schemes. 
Groth16 exhibits lower indicators overall when comparing the same proof with a different scheme, with the exception of the witness computation. 
On average, comparing Groth16 with Marlin, there is an increase of 71\% for Gas, 40930\% for setup and 1200\% for proof times, 16.75\% for compiled ecs size, 3698\% for private key size, 12.5\% for verification key size and a decrease of 20\% for witness computation time. 

To assess the practical implications for DePINs, we can distinguish between \textit{one-time and recurring operations}. 
The setup is executed once per proof specification resulting in one zkVPR. 
A single zkVPR may be sufficient for a DePIN project if it contains all necessary registration conditions. 
However, when the device registration conditions change, a new zkVPR must be created resulting in another setup execution.
While the setup using Marlin leads to long execution times, such overheads can be considered insignificant in practice as zkVPRs are not expected to change frequently. 

\textit{Recurring operations}, i.e., witness computation, proof generation, and verification, are executed once per device registration. 
The combined off-chain operations of witness computation and proof generation do not exceed 11 seconds for Groth16 but with Marlin they lead up to 83 seconds for membership proofs. 
Similarly, transaction costs increase when using Marlin. 
While both, off-chain execution times and on-chain transaction costs, are bearable for single devices registration even for Marlin, they may represent a hurdle for time-critical registration of large numbers of devices. 


        \section{Discussion}
	\label{sec:discussion}
In this Section, we revisit security objectives from Section~\ref{sec:04_model} and discuss yet unaddressed issues of the system.

\subsection{Trusted Setup}
By using Marlin~\cite{chiesa2020marlin} as a proof system, we can reduce trust assumptions from the setup and protect against attacks of the dApp initiator. 
With Marlin, the $srs$ is updated over time by multiple parties with the security assumption that at least one participant in each update phase is honest. 
Each update enhances the trustworthiness of the $srs$ because it adds layers of contributions from various parties.
In the proposed system, device owners can participate in the updating procedure to be certain about the trustworthiness. 
However, as our experiments show, these improved security guarantees come at the cost of higher transaction costs and execution times. 

Beyond that, such attacks can be prevented through an adjustment of the setup phase, e.g., employing a blockchain-based setup ceremony as proposed in~\cite{park2020smart} or leveraging secure multi-party computation (sMPC) as proposed in~\cite{ben2015secure}.
Furthermore, different proof protocols could be applied that do not require a trusted setup like STARKs~\cite{ben2018scalable} which, however, are more expensive to verify.
As another approach, some zero-knowledge Virtual Machines (zkVM) like Risc Zero combine STARKs and SNARKS to prove the correct execution of the VM's instruction. 
In contrast to a per-application setup for SNARKs, zkVMs rely on a single setup for the zkVM that is often well documented\footnote{\url{https://dev.risczero.com/api/trusted-setup-ceremony}}. 

\subsection{Replay Attacks}
Replay attacks pose another significant problem for the system.
After on-chain verification, proofs become publicly accessible, allowing anyone on the blockchain to resubmit already accepted zkVPs to impersonate another identity and gain unauthorized registration. 
To mitigate this issue, uniqueness proofs can be used as proposed in~\cite{heiss2022non} which enable the dApp to detect repeated submissions of the same proof by different users. 

\subsection{Revocation}
The issuer may need to revoke a credential or a credential may be subject to change, e.g., if a device's firmware is updated, the version number changes. 
Although revocation is not covered in this work, we suggest using specialized blockchain-based revocation systems as an extension to our credential on-chaining system, similar to the methods proposed for educational credentials in~\cite{vidal2020revocation}. 
Additionally, in some instances, revocation can be substituted with expiration dates on credentials, which can be implemented using relative time-dependent proofs as proposed in~\cite{heiss2022non}.

\subsection{Public Key Disclosure}
The device registry of the registration contract contains the public keys of all registered devices ($pubk_d$). 
The disclosure of the $pubk_d$ on the blockchain opens the opportunity for attacks on the device, e.g., key search attacks. 
To protect the $pubk_d$, we propose the following extension: 
Only a hash-based commitment to $pubk_d$ is stored in the registry. 
To authenticate, the device owner creates a zero-knowledge proof that encodes (1) the verification of the signature using $pubk_d$ as private input and (2) proves that the $pubk_d$ belongs to the commitment which is treated as public inputs. 
On-chain, the correctness can be verified through a successful proof verification and a comparison of the public input with the corresponding $pubk_d$ commitment of the registry.

	\section{Related Work}
	\label{sec:relatedWork}
	To our best knowledge, we provide the first credential-based device registration procedure for dApps and DePINs.
However, our research intersects with various studies on the applicability of self-sovereign identity (SSI) concepts in the Internet of Things (IoT), the use of verifiable credentials (VC) in blockchain-based decentralized applications (dApps), and zero-knowledge proof (ZKP)-based anonymous credentials.

\textbf{SSI for IoT}:
The application of self-sovereign identity for IoT devices has been examined in several studies.
Fedrecheski et al.~\cite{fedrecheski2020self} compare existing models for device identities, attributes, and key management, including PGP, X.509, and SSI.
Dayaratne et al.~\cite{dayaratne2024ssi4iot} elaborate on the potential of SSI for IoT through a taxonomy that helps compare different device issuers, IoT use cases, and device lifecycle management.
Gebresilassie et al. propose and evaluate an SSI-based system for IoT in the context of a car rental use case~\cite{gebresilassie2020distributed}.
Additional use cases for SSI in industrial IoT are discussed by another study~\cite{8869262}.
In these approaches, blockchain technology is used as a Verifiable Data Registry (VPR) rather than as an application platform. In contrast, our work focuses on smart contract-based validation of device credentials, enabling their integration in dApps.

\textbf{SSI in dApps}:
Several studies address the use of device credentials in dApps.
The DIAM-IoT framework, introduced by Fan et al.~\cite{fan2020diam}, is a decentralized Identity and Access Management (IAM) system designed for data sharing in the IoT.
Luecking et al.~\cite{luecking2020decentralized} propose an SSI-based system for IoT devices that uses blockchain technology to host a reputation system for these devices.
Another study~\cite{9621153} utilizes decentralized identifiers and VCs in blockchain-based data trading systems to authenticate users and prove data ownership.
While these works extend the use of blockchains beyond VDR functionality, they do not provide smart contract-based validation of device credentials.

\textbf{Anonymous Credentials in dApps}:
ZKP-enabled anonymous credentials are explored for off-chain verifiers in various industrial projects like Privado (former PolygonID)~\cite{misc:privado} or Hyperledger AnonCreds~\cite{misc:HLIndy}. 
Research on smart contract-enabled on-chain verifiers is conducted only in a few proposals.
Yin et al.~\cite{yin2022smartdid} propose an identity system for IoT based on consortium blockchains that use commitments and ZKPs to protect confidential attributes on-chain. While this system supports on-chain credential verification, it is not suitable for integrating verification results in dApps without trusted intermediaries.
Muth et al.~\cite{muth2022towards} demonstrate how CL-signature-based anonymous credentials from the Hyperledger Indy ecosystem can be verified by smart contracts running in the Ethereum Virtual Machine.
Heiss et al.~\cite{heiss2022non} continue this work by applying zkSNARK-based verifiable off-chain computation to enable privacy-preserving credential verification in dApps.
These approaches are closely related as they verify credentials on the blockchain, but they do not focus on device credentials, which is the primary focus of our research.

	\section{Conclusion}
	\label{sec:conclusion}
	In this paper, we introduced a credential-based device registration (CDR) mechanism for dApps in DePINs. 
Our mechanism uses zkSNARKs to allow device owners to create zero-knowledge verifiable presentations (zkVPs) based on requests (zkVPRs) from dApp initiators. 
These zkVPs can be verified on the blockchain without revealing confidential device attributes, ensuring transparent verification by other device owners.
We provided a detailed characterization of CDRs, including registration conditions, device attributes, and credential issuers, and presented a system model integrating CDRs with the W3C VC model. We implemented the CDR using ZoKrates~\cite{eberhardt2018zokrates} to create and verify zkSNARKs-based zkVPRs and zkVPs on the Ethereum blockchain. 
Our evaluation with Groth16~\cite{groth2016size} and Marlin~\cite{chiesa2020marlin} on different registration conditions gave first insights into the performance impact of applied proof systems and highlighted a tradeoff between security and efficiency.
While we see CDRs as a crucial mechanism to establish trust in DePINs, we consider this work preliminary and plan to continue our research. 
In addition to the discussed security concerns, future work will address the integrability with existing DePIN technologies like W3bstream, the applicability of alternative proving environments like zero-knowledge virtual machines, architectural variants considering self-sovereign devices, and further experimentation, for example, using constraint hardware like smartphones for proving.

	\bibliographystyle{IEEEtran}
	\bibliography{./references}

	\vspace{12pt}

\end{document}